\documentclass[a4paper,fleqn,usenatbib]{mnras}
\usepackage{newtxtext,newtxmath}
\usepackage[T1]{fontenc}
\usepackage{ae,aecompl}
\usepackage{amsmath,color,xspace,multirow,longtable,graphicx,bigints,comment}
\usepackage{tabularx}

\newcommand\clearrow{\global\let\rowmac\relax}
\clearrow

\newcommand{\dc}{DC\,818760\xspace}
\newcommand\textlcsc[1]{\textsc{\MakeLowercase{#1}}}
\newcommand{\cii}{[C\,II]\xspace}

\title[A Dusty Triple Merger at $z\sim4.56$]{The ALPINE-ALMA [CII] Survey: A Triple Merger at $\mathbf{z\sim4.56}$}
\author[G. C. Jones et al.]{G. C. Jones$^{1,2}$\thanks{E-mail: gj283@cam.ac.uk},
M. B\'{e}thermin$^{3}$, 
Y. Fudamoto$^{4}$, 
M. Ginolfi$^{4}$,
P. Capak$^{5}$, 
P. Cassata$^{6}$,\newauthor
A. Faisst$^{5}$,  
O. Le F\`evre$^{3}$, 
D. Schaerer$^{4}$, 
J. Silverman$^{7}$, 
L. Yan$^{5,8}$,
S. Bardelli$^{9}$,\newauthor
M. Boquien$^{10}$,
A. Cimatti$^{11,12}$,
M. Dessauges-Zavadsky$^{4}$,
M. Giavalisco$^{13}$,
C. Gruppioni$^{9}$, \newauthor
E. Ibar$^{14}$,
Y. Khusanova$^{3}$,
A. M. Koekemoer$^{15}$,
B. C. Lemaux$^{16}$,
F. Loiacono $^{9,11}$,\newauthor
R. Maiolino$^{1,2}$,
P. A. Oesch$^{4,17}$, 
F. Pozzi$^{9}$,
D. Riechers$^{18,19}$\thanks{Humboldt Research Fellow},
G. Rodighiero $^{6}$,\newauthor
M. Talia$^{9,11}$,
L. Vallini$^{20}$,
D. Vergani$^{9}$,
G. Zamorani$^{9}$,
E. Zucca$^{9}$
\\
$^{1}$Cavendish Laboratory, University of Cambridge, 19 J. J. Thomson Ave., Cambridge CB3 0HE, UK\\
$^{2}$Kavli Institute for Cosmology, University of Cambridge, Madingley Road, Cambridge CB3 0HA, UK\\
$^{3}$Aix Marseille Univ, CNRS, CNES, LAM, Marseille, France\\
$^{4}$Observatoire de Gen\`eve, Universit\'e de Gen\`eve  51 Ch. des Maillettes, 1290 Versoix, Switzerland \\
$^{5}$IPAC, California Institute of Technology,  1200 East California Boulevard, Pasadena, CA 91125, USA \\
$^{6}$University of Padova, Department of Physics and Astronomy  Vicolo Osservatorio 3, 35122, Padova, Italy\\
$^{7}$Kavli Institute for the Physics and Mathematics of the Universe, The University of Tokyo  Kashiwa, Chiba 277-8583, Japan \\
$^{8}$Caltech Optical Observatories, Cahill Center for Astronomy and Astrophysics  1200 East California Boulevard, Pasadena, CA 91125, USA\\
$^{9}$Osservatorio di Astrofisica e Scienza dello Spazio - Istituto Nazionale di Astrofisica, via Gobetti 93/3, I-40129, Bologna, Italy\\
$^{10}$Centro de Astronom\'ia (CITEVA), Universidad de Antofagasta, Avenida Angamos 601, Antofagasta, Chile\\
$^{11}$University of Bologna, Department of Physics and Astronomy (DIFA), Via Gobetti 93/2, I-40129, Bologna, Italy\\
$^{12}$INAF - Osservatorio Astrofisico di Arcetri, Largo E. Fermi 5, I-50125, Firenze, Italy\\
$^{13}$ Department of Physics and Astronomy, University of Massachusetts, Amherst, MA 01003, USA\\
$^{14}$ Instituto de F\'isica y Astronom\'ia, Universidad de Valpara\'iso, Avda. Gran Breta\~na 1111, Valpara\'iso, Chile\\
$^{15}$Space Telescope Science Institute, 3700 San Martin Dr., Baltimore, MD 21218, USA\\
$^{16}$Department of Physics, University of California, Davis, One Shields Ave., Davis, CA 95616, USA\\
$^{17}$International Associate, Cosmic Dawn Center (DAWN) at the Niels Bohr Institute, University of Copenhagen and DTU-Space, Technical University of Denmark\\
$^{18}$Department of Astronomy, Cornell University, Space Sciences Building, Ithaca, NY 14853, USA\\
$^{19}$Max-Planck-Institut f\"ur Astronomie, K\"onigstuhl 17, D-69117 Heidelberg, Germany\\
$^{20}$Leiden Observatory, Leiden University, PO Box 9500, 2300 RA Leiden, The Netherlands
}

\begin{document}
\label{firstpage}
\pagerange{\pageref{firstpage}--\pageref{lastpage}}
\maketitle

\begin{abstract}
We report the detection of [CII]$\lambda158{\rm \mu m}$ emission from a system of three closely-separated sources in the COSMOS field at $z\sim4.56$ , as part of the ALMA Large Program to INvestigate CII at Early times (ALPINE). 
The two dominant sources are closely associated, both spatially ($1.6''\sim11$\,kpc) and in velocity ($\sim100$\,km\,s$^{-1}$), while the third source is slightly more distant ($2.8''\sim18$kpc, $\sim300$\,km\,s$^{-1}$). The second strongest source features a slight velocity gradient, while no significant velocity gradient is seen in the other two sources. 
Using the observed \cii luminosities, we derive a total log$_{10}(\rm SFR_{[CII]}\,[M_{\odot}\,year^{-1}])=2.8\pm0.2$, which may be split into contributions of 59\%, 31\%, and 10\% from the central, east, and west sources, respectively.
Comparison of these \cii detections to recent zoom-in cosmological simulations suggests an ongoing major merger. We are thus witnessing a system in a major phase of mass build-up by merging, including an on-going major merger and an upcoming minor merger, which is expected to end up in a single massive galaxy by $z\sim2.5$.
\end{abstract}

\section{Introduction} 
Cosmological zoom-in simulations of high-redshift galaxies (i.e., $z>4$) show that they built up mass through a complex process, with both continuous gas accretion from diffuse haloes and discrete episodes of major and minor mergers (e.g., \citealt{vall13,pall17,koha19,pall19}). 

While secular accretion is difficult to directly observe due to its low-excitation nature, observational evidence of merging at these high redshifts is pervasive. The brightest examples of merging are galaxies undergoing bursts in star formation apparently driven by major mergers (e.g., \citealt{oteo16,riec17,pave18,marr18}).
In addition, resolved spectral observations have revealed evidence of merging in star-forming main-sequence galaxies (SFGs; e.g., \citealt{noes07,spea14}), or galaxies whose stellar masses and star formation rate show a correlation, with a normalization that evolves with redshift. 
This evidence of merging is manifest as a clumpy morphology (\citealt{ouch13,riec14,maio15,capa15,will15,carn17,bari17,jone17,math17,ribi17,carn18a,carn18b,matt19}), which is interpreted as ongoing galaxy assembly via minor mergers. The presence of clumps may also be explained by gravitational instabilities inside disk galaxies (e.g., \citealt{ager09}), and the true nature of a source may only be revealed using detailed kinematic information (e.g., from spectroscopy).
Despite the number of individual detections, the sample of observed mergers at $z>4$ confirmed from dynamical arguments is still statistically low, and more detections are required in order to characterize the merger rate as a function of cosmological time at these high redshifts.

The need for a more systematic merger identification and characterization at $z>4$ can be fulfilled by the ALMA Large Program to INvestigate CII at Early times (ALPINE; \citealt{fais19}, Le F\`evre et al. in prep.), which observed \cii$\lambda158{\rm \mu m}$ emission and rest-frame $\sim158$\,$\mu$m continuum emission from 118 SFGs in the Cosmic Evolution Survey (COSMOS) and Extended Chandra Deep Field-South (ECDFS) fields with $4.4<z_{spec}<5.8$, SFR$>10$\,M$_{\odot}$\,year$^{-1}$, log(M$_*$/M$_{\odot})=9-11$, and L$_{UV}>0.6$\,L$^*$. These cuts were made to ensure that the sample represents the overall galaxy population at this epoch

Since \cii is generally the brightest FIR emission line for star forming galaxies \citep{cariw13} and is emitted from all the gas phases (ionized, neutral and molecular) of the interstellar medium (ISM; \citealt{pine13}), it is a prime tracer of the gas kinematics of high-redshift galaxies. As an example of mergers identified in the ALPINE survey, we detail here the detection of \cii and dust continuum emission from the $z_{\rm [CII]}=4.56$ dusty triple merger DEIMOS\_COSMOS\_818760 (hereafter \dc).

Because it is located in the well-studied COSMOS field (\citealt{scov07a,scov07b}), \dc has been observed with a number of NUV-NIR instruments, including HST, Subaru, and Spitzer \citep{laig16}. Using the broadband (i.e., CFHT \textit{u} through Spitzer IRAC 8.0\,$\mu$m) SED of \dc and the SED modelling code LePHARE (\citealt{arno99,ilbe06,arno11}) with a Chabrier initial mass function and Calzetti starburst extinction law, Faisst et al. (in prep) find a stellar mass of log(M$_{*}\,[M_{\odot}])=10.6\pm0.1$ and a star formation rate log(SFR\,$[M_{\odot}\,$year$^{-1}])=2.7^{+0.2}_{-0.3}$\,M$_{\odot}$\,year$^{-1}$. These values place \dc on the upper envelope of the main sequence at $z\sim4.6$ \citep{spea14,tasc15}. 

In addition, \dc is nearby (i.e., $\sim5.5$ proper Mpc and $<500$\,km\,s$^{-1}$) the massive protocluster PC1 J1001+0220 \citep{lema18}. Since it lies along the major axis of the protocluster, and is only $\sim3.5$ proper Mpc from the northeast component of this protocluster, \dc may be associated with the system in a filamentary structure.

In this work, we discuss new ALMA observations of \cii and submm continuum emission from \dc obtained as part of the ALMA large program ALPINE, and examine its triple merger nature.
We assume a flat $\Lambda$-CDM cosmology ($\Omega_{\Lambda}=0.7$, $\Omega_{m}=0.3$, H$_{o}=70$\,km\,s$^{-1}$) throughout. 
At the redshift of \dc ($z_{\rm [CII]}=4.560$), 1\,arcsecond corresponds to 6.563 proper kpc . 

\section{Observations \& Data Reduction}
The \cii emission from \dc was observed with ALMA on 25 May, 2018 in cycle 5 (project 2017.1.00428.L, PI O. 
Le F\`evre) using configuration C43-2 (baselines $\sim15-320$\,m), 45\,antennas, and  an on-source time of 17\,minutes. J0158+0133 was used as a bandpass and flux calibrator, while J0948+0022 was used as a phase calibrator.

The spectral setup consisted of two sidebands, each constructed of two spectral windows (SPWs) of width 1.875\,GHz. Each SPW was made of channels of width 15.625\,MHz. The lower sideband is tuned to the redshifted \cii frequency, while the upper sideband is solely used for continuum.

Calibration was performed using the heuristic-based CASA 5.4.1 automatic pipeline, with reduced automatic band-edge channel flagging (B\'{e}thermin et al. in prep). The pipeline calibration diagnostics were inspected carefully and no issues were found. To maximise sensitivity, we adopted natural weighting.

Continuum and line emission were separated using the CASA task \textlcsc{uvcontsub}. 
The lower sideband was made into a data cube using the CASA task \textlcsc{tclean}, resulting in an average RMS noise level per 15.625\,MHz ($\sim14$\,km\,s$^{-1}$) channel of 0.6\,mJy\,beam$^{-1}$ and a synthesized beam of $1.07''\times0.84''\,$at$\,-81^{\circ}$.
To maximize sensitivity, one continuum image was created using all line-free data in both sidebands using \textlcsc{tclean} in multi-frequency synthesis mode. This results in a continuum image with an RMS noise level of 0.05\,mJy\,beam$^{-1}$ and the same synthesized beam as the upper sideband.

\section{Imaging Results}

\begin{figure}
\centering
\includegraphics[width=\columnwidth]{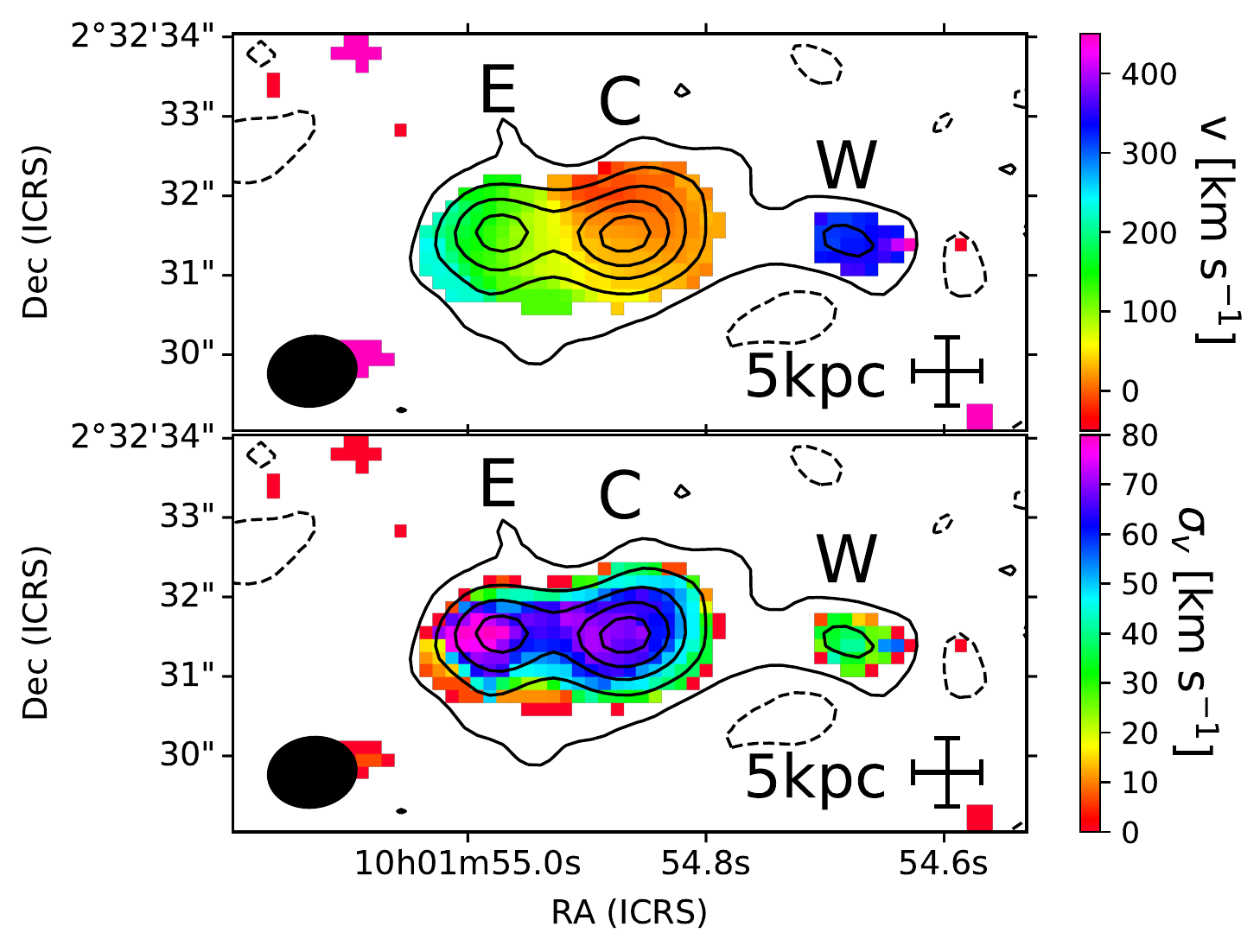}
\caption{Total \cii moment zero map (contours), velocity field (top color), and velocity dispersion map (bottom color). Contours begin at $\pm2\sigma$, where $1\sigma= 0.1\,$Jy\,beam$^{-1}$\,km\,s$^{-1}$, and are in steps of $5\sigma$. Zero velocity is defined as $z_{\rm [CII]}=4.560$, or the redshift of the central source. The synthesized beam ($1.07''\times0.84''$, with major axis position angle = $-82^{\circ}$) is shown by the solid black ellipse to the lower left. A 5\,kpc$\times$5\,kpc scale is shown in the lower right. North is up and east is to the left.}
\label{mom0}
\end{figure}

\begin{figure}
\centering
\includegraphics[width=\columnwidth]{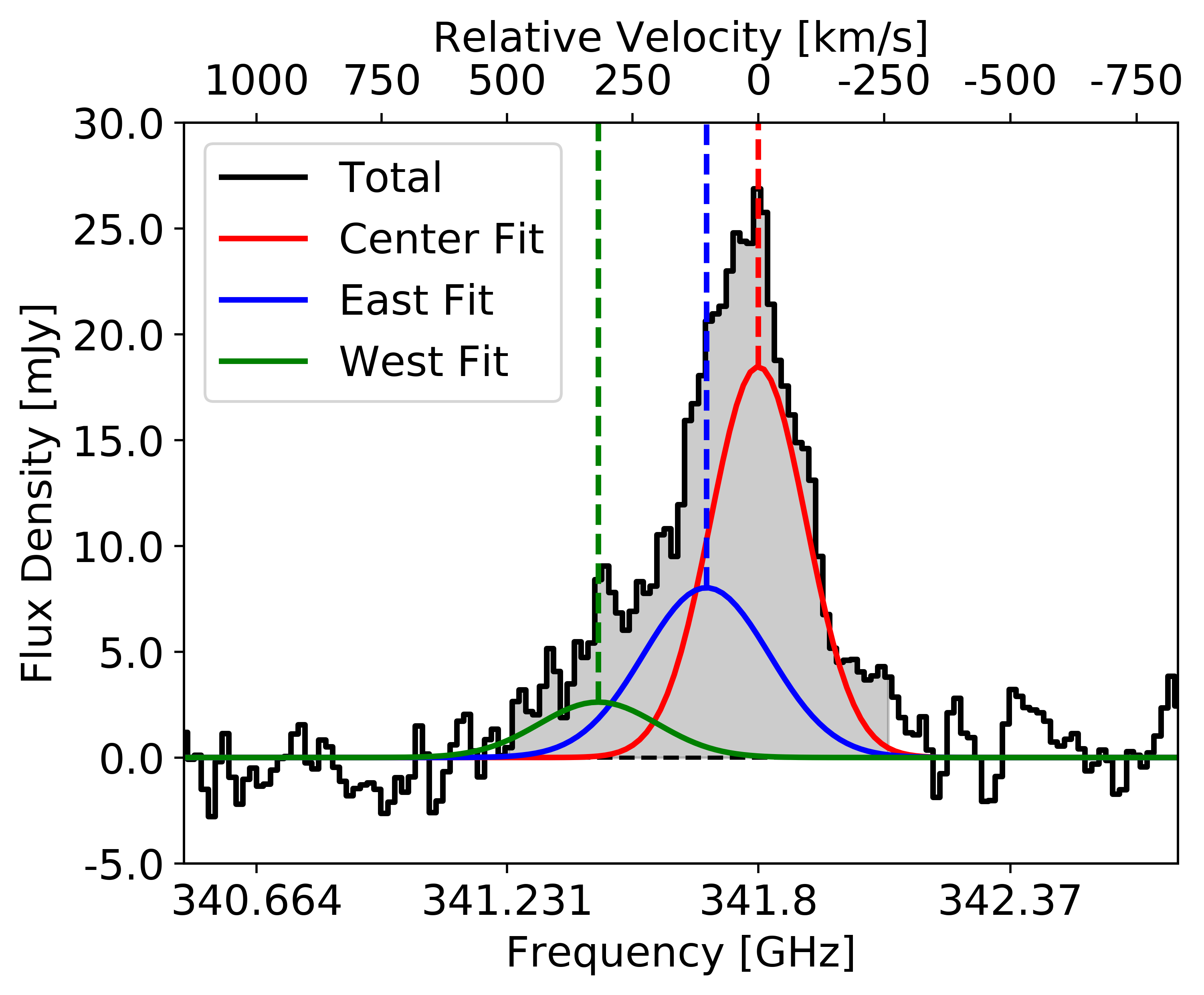}
\caption{
Global spectrum taken over $2\sigma$ contour of total \cii moment zero map (black histogram). 
Shaded region shows channels used to create moment zero image (contours of Figure \ref{mom0}).
An approximation of the contribution of each source is shown by solid colored lines, with the central frequency marked by a vertical dashed line of the same color.
}
\label{globspec}
\end{figure}

In order to investigate the \cii emission in this source, we first examine which channels show emission with $>2\sigma_{\rm LINE}$ ($\sigma_{\rm LINE}$=0.6\,mJy\,beam$^{-1}$). Using these channels (see shaded channel range of Figure \ref{globspec}) and the CASA task \textlcsc{immoments}, we create a moment zero image of the total \cii emission (contours of Figure \ref{mom0}). Three \cii sources are present, all roughly at the same declination.

Using the CASA task \textlcsc{imfit}, we simultaneously fit three two-dimensional Gaussians to the moment zero map. The central source of this best fit (C)  features an integrated flux density $(4.9\pm0.3)$\,Jy\,km\,s$^{-1}$, a peak flux density $(2.4\pm0.1)$\,Jy\,beam$^{-1}$\,km\,s$^{-1}$, and a beam-deconvolved size $(1.2\pm0.1)''\times(0.8\pm0.1)''$ at a position angle $(111\pm11)^{\circ}$, defined counterclockwise from north. The second source (E), which is $1.6''$ ($\sim11$\,kpc) to the east of source C, shows an integrated flux density $(2.6\pm0.2)$\,Jy\,km\,s$^{-1}$,  a peak flux density $(1.9\pm0.1)$\,Jy\,beam$^{-1}$\,km\,s$^{-1}$, and a deconvolved size $(0.8\pm0.1)''\times(0.3\pm0.2)''$ at a position angle $(9\pm169)^{\circ}$. Lastly, the weakest source, (W) which is $2.8''$ ($\sim18$\,kpc) west of source C, has an integrated flux density $(0.8\pm0.2)$\,Jy\,km\,s$^{-1}$, a peak flux density $(0.9\pm0.1)$\,Jy\,beam$^{-1}$\,km\,s$^{-1}$, and is unresolved. 

The kinematics of this field are revealed by the velocity field (moment one image), created using the CASA task \textlcsc{immoments} (top panel color of Figure \ref{mom0}). 
While the two brightest sources (i.e., C and E) are only separated by $\sim100$\,km\,s$^{-1}$, source W is $\sim300$\,km\,s$^{-1}$ offset from source C. Sources C and W show nearly constant velocity, while source E shows a strong gradient ($50\sim200$\,km\,s$^{-1}$).

The \textlcsc{immoments} task may also be used to create a velocity dispersion (moment two) map (bottom panel color of Figure \ref{mom0}). Source W shows a relatively low velocity dispersion ($\sigma_{v,W,pk}\sim40$\,km\,s$^{-1}$), while sources C and E exhibit strong peaks in velocity dispersion ($\sigma_{v,C,pk}\sim70$\,km\,s$^{-1}$, $\sigma_{v,E,pk}\sim80$\,km\,s$^{-1}$). These $\sigma_v$ peaks may be artificially enhanced by beam smearing (e.g., \citealt{wein06}), but each is spatially coincident with a \cii source.

Extracting a spectrum over the $2\sigma$ contour of the moment zero map, which contains all three sources, we obtain the profile shown in Figure \ref{globspec} (black line). In order to determine the contribution of each source, we first assign each spaxel within the $2\sigma$ moment zero contour to one of the three sources, based on the relative contributions of each of the three Gaussian components output from CASA \textlcsc{imfit}. The corresponding pixels for each source are then integrated to produce three integrated spectra. For clarity, each spectrum is fit with a one-dimensional Gaussian, and displayed in Figure \ref{globspec}, along with its centroid velocity. We find that sources E and W are both redshifted with respect to source C by $\sim100$ and $\sim300$\,km\,s$^{-1}$, respectively. The redshift of the dominant source C, which will be used as the redshift of this field, is $4.56038\pm0.00004$.

\begin{figure}
\centering
\includegraphics[width=\columnwidth]{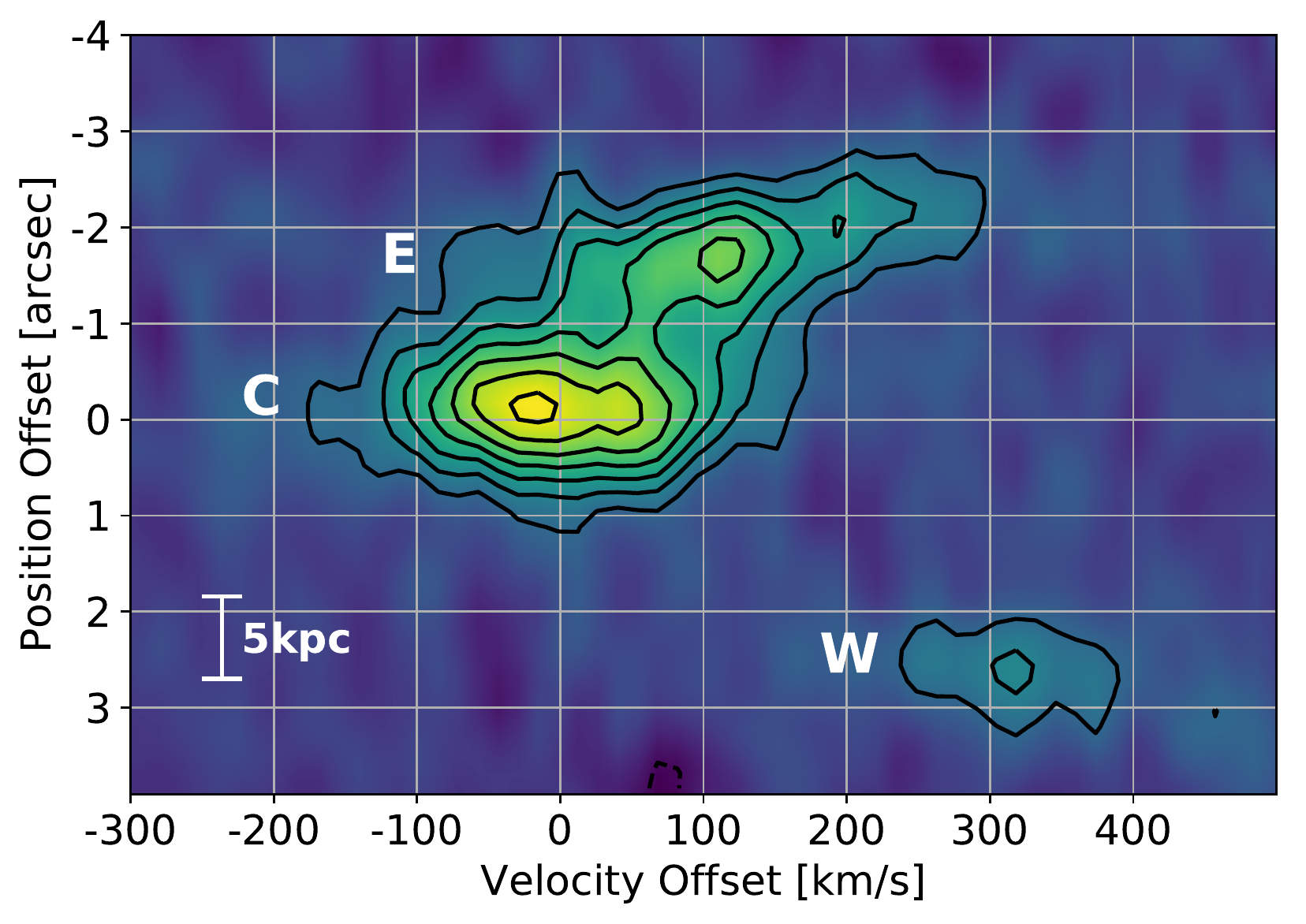}
\caption{Position-velocity diagram taken east-west, centered on the central source, with a total width of $8''$ and an averaging width of five pixels. Contours begin at $3\sigma$, where $1\sigma= 0.6\,$mJy\,beam$^{-1}$, and are in steps of $2\sigma$. East is up and west is down. Scale of 5\,kpc shown to lower left.}
\label{pv}
\end{figure}

To examine the kinematics of this system in another way, we create a position-velocity (PV) diagram (CASA \textlcsc{impv}) by extracting a 5 pixel thick (1\,pixel=$0.16''$), $8''$ long slice across the right ascension axis of the data cube, centered on the central source. This 3D slice was then averaged in declination to create a single intensity plane (see Figure \ref{pv}). This PV diagram confirms that both of the fainter sources (i.e., E and W) are moving at positive velocities with respect to source C, as seen in the spectra (Figure \ref{globspec}). While the sources C and E are closely connected, the source C and W are separated by $\sim18$\,kpc and $\sim300\,$km\,s$^{-1}$ . 

\begin{figure}
\centering
\includegraphics[width=\columnwidth]{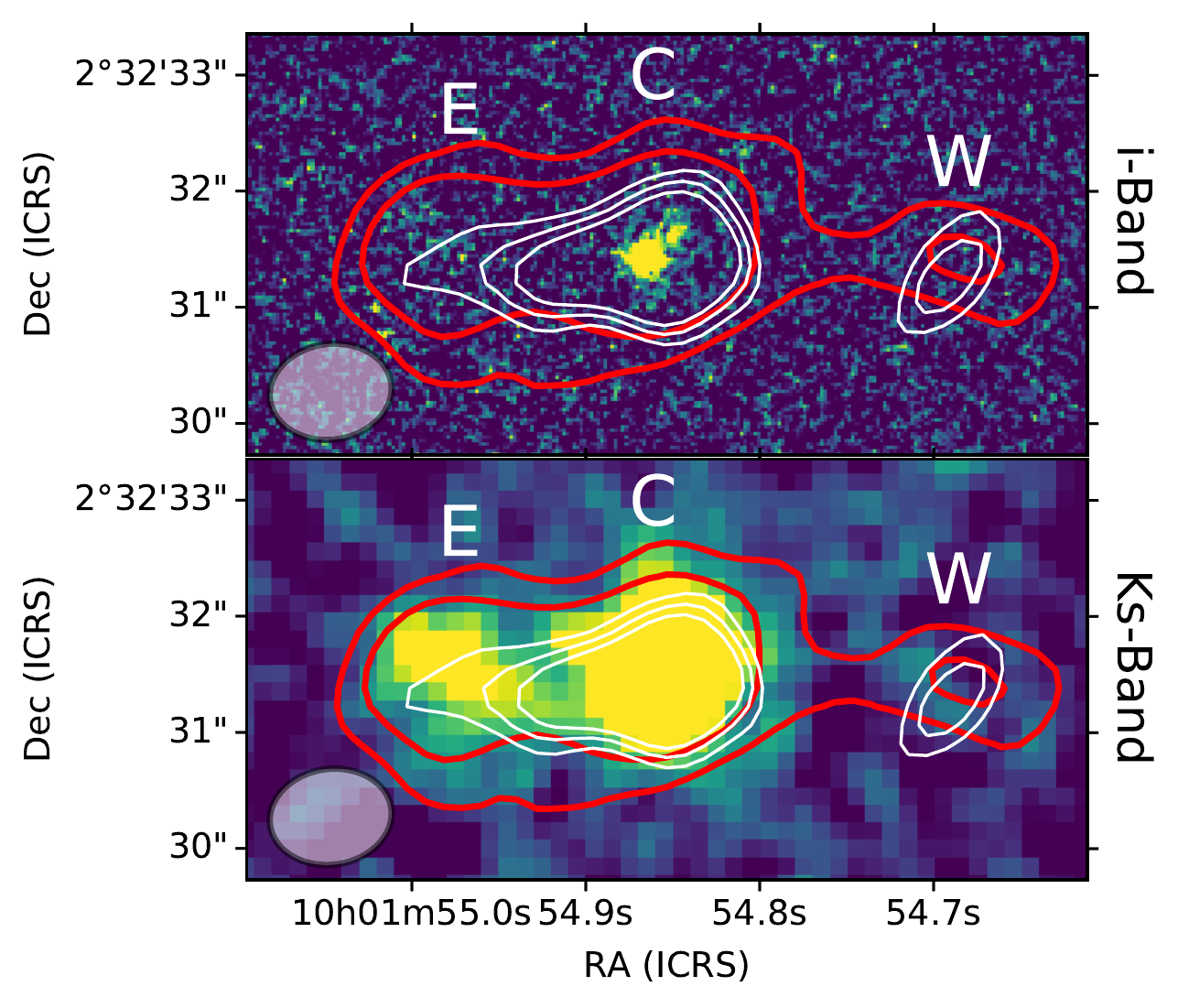}
\caption{
Background image of the top panel shows HST/ACS F814W (I-band) image of \dc (\citealt{koek07,scov07b}), while the bottom panel shows the UltraVISTA Ks image \citep{mccr12}. In both panels, white contours depict the continuum image created using all line-free data. Contours shown at $3,4,5\sigma$, where $1\sigma= 0.05\,$mJy\,beam$^{-1}$. The synthesized beam ($1.06''\times0.82''$, with major axis position angle = $-82^{\circ}$) is shown by the solid black ellipse to the lower left. For reference, the (3,7)$\sigma$ \cii moment zero contours are shown in red.  North is up and east is to the left.
}
\label{cont}
\end{figure}

In addition to \cii emission, dust continuum emission is detected at the location of all three sources (white contours of Figure \ref{cont}). While the \cii emission features three distinct peaks, the bulk of the continuum emission is concentrated in source C, with a strong extension to the east, and a $4\sigma$ component coincident with source W. A single component, two-dimensional Gaussian fit to the combined emission of the source C and its eastern extension using CASA \textlcsc{imfit} yields a peak flux density of $0.44\pm0.05$\,mJy\,beam$^{-1}$, an integrated flux density of $1.22\pm0.18$\,mJy, and a deconvolved size of $(1.95\pm0.35)''\times(0.71\pm0.18)''$ at $(90\pm7)^{\circ}$. The continuum emission from source W is unresolved, with a peak flux density of $0.18\pm0.05$\,mJy\,beam$^{-1}$ and integrated flux density of $0.26\pm0.11$\,mJy. 

The HST/ACS F814W (i-band) image of this source reveals emission from only source C. This emission may be decomposed into two components separated by only $\sim0.3''$ (1.9\,kpc), possibly indicating a small-separation merger. On the other hand, the UltraVISTA Ks-band image of this field (which is $\sim2$\,magnitudes less sensitive) shows emission from both source C and E. Neither image shows significant emission from source W. This dramatic increase in the emission between the i-band ($\lambda_{\rm rest}\sim1450$\AA) and Ks-band ($\lambda_{\rm rest}\sim4000$\AA) for source E may indicate a steep UV slope, implying a significant dust presence (e.g., \citealt{calz00}). 

\section{Analysis}
\subsection{Star Formation Rate}

The empirical L$_{\rm [CII]}$ to SFR calibration of \citet{delo14} for their full sample of star-forming galaxies may be used to estimate the SFR of each source individually. We note that \citet{carn18b} found that a sample of $z>5$ \cii star-forming galaxies featured a $\sim2\times$ larger dispersion in this relation than is stated in \citet{delo14} for local galaxies, so the resulting uncertainties in SFR for \dc are likely slightly underestimated.
The three L$_{\rm [CII]}$-derived star formation rates (see Table \ref{tab:my_label}) sum to log$_{10}\rm (SFR_{[CII]}\,[M_{\odot}year^{-1}])=2.82\pm0.23$.  Comparing the different sources, we find that the star formation activity of the system may be split into contributions from source C ($59\%$), E ($31\%$), and W ($10\%$).

Using flux densities extracted from an aperture centered on source C of diameter $3''$, which encloses only source C and a portion of source E, Faisst et al. (in prep) created a broadband SED of \dc, and fit it with LePHARE. The resulting value of log$_{10}\rm (SFR_{SED}\,[M_{\odot}year^{-1}])=2.7^{+0.2}_{-0.3}$ is in agreement with our value for the source C of log$_{10}\rm (SFR_{[CII],C}\,[M_{\odot}year^{-1}])=2.59\pm0.24$, suggesting that [CII] is an appropriate SFR tracer in this source.

\begin{table*}
\centering
\begin{tabular}{c|ccc|cc}
 										 		   & C 					 		& E 				   		 & W 					 	  & CE 		      & CEW\\ \hline
Right Ascension									   & 10h01m54.865s				& 10h01m54.978s 			 & 10h01m54.683s  			  & $\cdots$       & $\cdots$\\
Declination										   & +2$^{\circ}$32$'$31.53$''$ & +2$^{\circ}$32$'$31.50$''$ & +2$^{\circ}$32$'$31.44$''$ & $\cdots$       & $\cdots$\\
$z_{\rm [CII]}$ 						 		   & $4.56038\pm0.00004$ 		& $4.56229\pm0.00008$ 		 & $4.56628\pm0.00014$ 		  & $\cdots$       & $\cdots$\\
$S\Delta$v$_{[CII]}$\,[Jy\,km\,s$^{-1}$] 		   & $4.9\pm0.3$		 		& $2.6\pm0.2$ 		   		 & $0.8\pm0.2$ 		 		  & $7.5\pm0.4$    & $8.3\pm0.4$\\
S$_{\rm cont}$\,[mJy] 					 		   & $\cdots$ 			 		& $\cdots$ 		   			 & $0.26\pm0.11$ 		 	  & $1.22\pm0.18$  & $1.48\pm0.21$\\ \hline
log(L$_{\rm [CII]}$\,[L$_{\odot}$]) 	 		   & $9.48\pm0.03$ 		 		& $9.21\pm0.03$ 	  	     & $8.70\pm0.11$ 		 	  & $9.67\pm0.02$  & $9.7\pm0.02$\\
log(SFR$_{\rm [CII]}$\,[M$_{\odot}$\,year$^{-1}$]) & $2.59\pm0.24$ 		 		& $2.31\pm0.23$ 	   		 & $1.80\pm0.25$ 		 	  & $2.77\pm0.24$  & $2.82\pm0.23$\\
\end{tabular}
\caption{Observed and derived quantities for each source in \dc, the combined quantities of the central and eastern sources, and the combined quantities of all three sources. 
Positional uncertainty is $\sim0.15''$.
SFR$_{[CII]}$ is derived from L$_{[CII]}$ using the full SFG relation of \citet{delo14}.
}
\label{tab:my_label}
\end{table*}

\subsection{Comparison to Simulations}

The recent zoom-in cosmological simulations of \citet{koha19} detail the evolution of a disk galaxy (``Alth\ae a'') undergoing minor and major merger events between $z=7.21-6.09$, and give both face-on and edge-on spectra for several evolutionary stages. In order to further characterize \dc, we compare our \cii observations with the results of these simulations. 

First, the global spectrum of \dc (Figure \ref{globspec}) shows an asymmetric Gaussian, composed of the dominant source C and the slightly weaker source E. This feature is also seen in the face-on merger spectrum of Alth\ae a (figure 6 of \citealt{koha19}). This similarity supports the merger interpretation of these two sources.

In addition, \citet{koha19} highlight the fact that galaxies with narrow spectral profiles (i.e., face-on disks, dispersion-dominated systems) are more easily observed than galaxies whose emission is spread over a broad velocity range (i.e., edge-on disks, mergers), due to their high peak flux density. This implies that additional components of the \dc system may also be present, but are too faint to be detected in our current observation. Indeed, the simulated galaxy Dahlia ($z\sim6$, SFR$\sim100$\,M$_{\odot}$\,year$^{-1}$; \citealt{pall17}) features 14 satellite clumps within 100\,kpc of its central galaxy, but only three of them were detected in \cii emission in a simulated observation. Thus, the \dc system is likely more complex than the three sources that we observe.

\section{Conclusions}
In this letter, we have presented the detection of \cii emission from three sources in the field \dc, as observed with ALMA as part of the large program ALPINE. The two dominant sources (C and E) are closely associated, both spatially ($1.6''\sim11$\,kpc) and in velocity ($\sim100$\,km\,s$^{-1}$), while the third source (W) is separate ($2.8''\sim18$kpc,$\sim300$\,km\,s$^{-1}$). Source E features a strong velocity gradient, which may either suggest a rotating galaxy or a tidally disrupted galaxy, while the others exhibit nearly constant velocity. All three show velocity dispersion peaks coincident with the peak of [CII] emission. 
Due to their kinematical properties, we conclude that the three sources in this field are separate objects, not members of the same galaxy. 

The close spatial separation, low velocity offset, and similar \cii luminosities (L$_{\rm [CII],C}/$L$_{\rm [CII],E}=1.86<4$) of sources C and E suggest an ongoing major merger \citep{lotz11}. 
Dynamical arguments from merger simulations (e.g. \citealt{kitz08}) indicate that these two sources will merge within $<0.5$\,Gyr. On the other hand, sources C and W are spatially and kinematically separate, but only by $\sim18$\,kpc and $\sim300$\,km\,s$^{-1}$. This close separation and their large luminosity ratio (L$_{\rm [CII],C}/$L$_{\rm [CII],W}=6.03>4$) suggests that they will coalesce in a minor merger at a later time. 

Based on both rest-frame UV observations and FIR continuum detections, there is strong evidence for significant internal extinction. Regarding the former, \dc was originally targeted with DEIMOS \citep{hasi18} using a slit coincident only with the central source. Ly$\alpha$ emission was not detected, but rather only UV ISM absorption lines, which strengthens the argument for dust obscuration.

Using the observed \cii luminosities, we derive a total log$_{10}\rm (SFR\,[M_{\odot}\,year^{-1}])=2.8\pm0.2$, which may be split into 59\%, 31\%, and 10\% from sources C, E, and W, respectively.
Comparison to cosmological zoom-in simulations show that the two dominant components resemble a merger and that the field likely contains multiple undetected sources

We are thus witnessing a massive galaxy in the early phase of mass assembly with merging playing a major role.
This system contains three kinematically distinct components: two currently undergoing a major merger (i.e., C and E), and a third minor component that will likely merge with the other two in the future (i.e., W).
 While the example given by this system is striking, larger samples are needed in order to assess how frequent such systems may be. The ALPINE sample is providing the opportunity to acquire a robust statistical knowledge of normal star-forming galaxies undergoing rapid mass growth, as will be presented in forthcoming papers.

\section*{Acknowledgements}
This paper is based on data obtained with the ALMA Observatory, under Large Program 2017.1.00428.L. ALMA is a partnership of ESO (representing its member states), NSF (USA) and NINS (Japan), together with NRC (Canada), MOST and ASIAA (Taiwan), and KASI (Republic of Korea), in cooperation with the Republic of Chile. The Joint ALMA Observatory is operated by ESO, AUI/NRAO and NAOJ. Based on data products from observations made with ESO Telescopes at the La Silla Paranal Observatory under ESO programme ID 179.A-2005 and on data products produced by TERAPIX and the Cambridge Astronomy Survey Unit on behalf of the UltraVISTA consortium.
This program is supported by the national program Cosmology and Galaxies from the CNRS in France. 
G.C.J. and R.M. acknowledge ERC Advanced Grant 695671 ``QUENCH'' and support by the Science and Technology Facilities Council (STFC). 
D.R. acknowledges support from the National Science Foundation under grant number AST-1614213 and from the Alexander von Humboldt Foundation through a Humboldt Research Fellowship for Experienced Researchers. 
A.C., F.P. and M.T. acknowledge the grant MIUR PRIN2017. 
L.V. acknowledges funding from the European Union’s Horizon 2020 research and innovation program under the Marie Sklodowska-Curie Grant agreement No. 746119. 
E.I.\ acknowledges partial support from FONDECYT through grant N$^\circ$\,1171710.

\bibliographystyle{mnras}
\bibliography{dc}

\label{lastpage}
\end{document}